**Title:** (Review) Developments in Multi-Chain Coarse-Grained Models for Entangled Polymer Dynamics

**Author:** Yuichi Masubuchi

Department of Materials Physics, Nagoya University, Nagoya 464-8603, Japan

**TOC graphics**

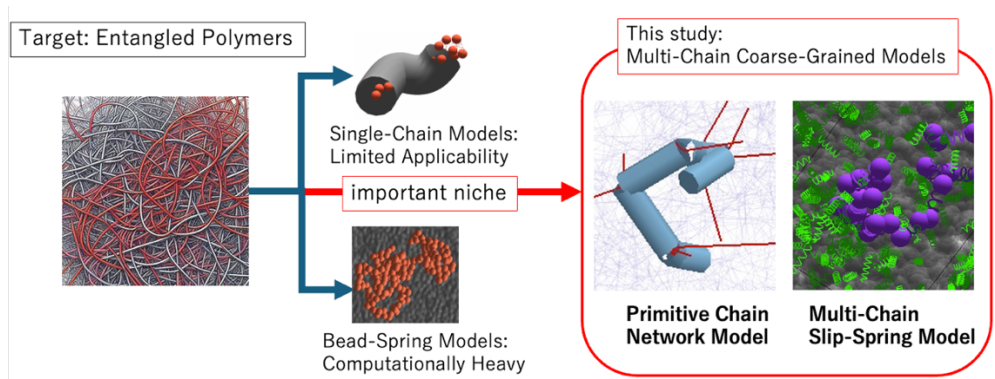

**Abstract**

This review describes the development and applications of multi-chain coarse-grained simulations for entangled polymer dynamics. The mean-field tube model has long served as the standard paradigm for describing the many-body entanglement problem as the motion of a single chain in a static field; it faces intrinsic limitations when addressing spatial correlations, fluctuations, and complex topological rearrangements. To overcome these limitations, "multi-chain" approaches—specifically the primitive chain network and multi-chain slip-spring models—were developed. These simulations explicitly resolve the force balance and topological coupling between multiple chains in three-dimensional space. This review covers the primitive chain network model, which emphasizes real-space force balance, and the multi-chain slip-spring model, which is derived from a well-defined free-energy functional. Linear and nonlinear rheology predictions are discussed, along with molecular mechanisms such as constraint release and stretch/orientation-induced reductions in friction. Extensions to branched polymers, wall-slip phenomena, and network polymers are also mentioned.

# 1. Introduction

Entangled polymer dynamics is a central topic in soft matter physics[1–6]. The viscoelastic behavior of these materials—elastic at short timescales and viscous at long timescales—arises from topological constraints due to the mutual uncrossability of long chains. The tube model by de Gennes[7], Doi, and Edwards[8–10] has been the most successful theoretical framework for these dynamics. In the tube model, a single chain is assumed to be confined within a tube-like potential formed by its neighbors, reducing the many-body problem to a tractable single-chain problem. This mean-field approach enabled the development of constitutive equations and computation algorisms that are now standard in polymer science[11–16].

The mean-field nature of the tube model, however, limits its applicability. The confining potential is typically treated as static or affinely deforming with the macroscopic flow. This approximation breaks down when multi-chain correlations, spatial heterogeneities, or thermodynamic fluctuations become important. In polydisperse blends, for example, the tube is formed by chains with different relaxation times; as the matrix chains relax, the tube constraint vanishes, a phenomenon known as constraint release (CR)[17–20]. Tube models have been refined to include CR and convective constraint release (CCR)[21,22], but these corrections are typically introduced as perturbative additions or ad hoc parameters such as the dynamic dilution exponent. In fast nonlinear flows, the assumption of affine tube deformation is often violated, and the microscopic origins of nonlinear rheological responses remain under debate.

To bridge the gap between macroscopic constitutive equations and atomistic or bead-spring molecular dynamics (MD) simulations, multi-chain coarse-grained models were developed[23–25]. These models operate at a mesoscopic level: coarse-grained enough to access experimental timescales (seconds to even hours) yet retaining the discrete, particle-like nature of polymer chains. The system is represented by an ensemble of chains interacting via discrete entanglements—modeled as slip-links or slip-springs—in three-dimensional space. This approach allows explicit simulation of entanglement creation and destruction, force balance at entanglement points, and network deformation without invoking a mean-field tube.

This review describes two such models developed by the author and colleagues: the primitive chain network (PCN) and the multi-chain slip-spring (MCSS) models. The PCN model uses 3D force balance among entangled subchains to describe network dynamics.

[26] It reproduces linear and nonlinear rheology by naturally capturing CR and CCR through interchain force interactions. However, the PCN model relies on phenomenological rules for topology change and an artificial osmotic potential to maintain density. The MCSS model was developed to address these issues. Its dynamics, including the hopping of slip-springs and their creation/destruction, are derived from a well-defined free energy functional under the principle of detailed balance[27].

In the following sections, the model constraints and governing equations are first described. The relationship between model parameters and the experimentally observable plateau modulus is then discussed, followed by the mapping to the standard Kremer-Grest bead-spring model[28]. Some applications to linear and nonlinear rheology are reviewed, and extensions to interfaces and gels are also mentioned.

## 2. Primitive Chain Network (PCN) Model

The PCN model represents entangled polymers as a three-dimensional network of "primitive chains" connected by slip-links, as shown in Fig 1. Unlike single-chain slip-link models[29–33], where the constraints are represented by a static field, the PCN constitutes a coupled network in real space. A polymer chain is described as a sequence of subchains (network strands) connecting consecutive slip-links (network nodes). The state variables of the system are the position vectors of the slip-links $\{\mathbf{R}\}$, the number of Kuhn segments in each subchain $n$, and the number of subchains per chain $Z$.

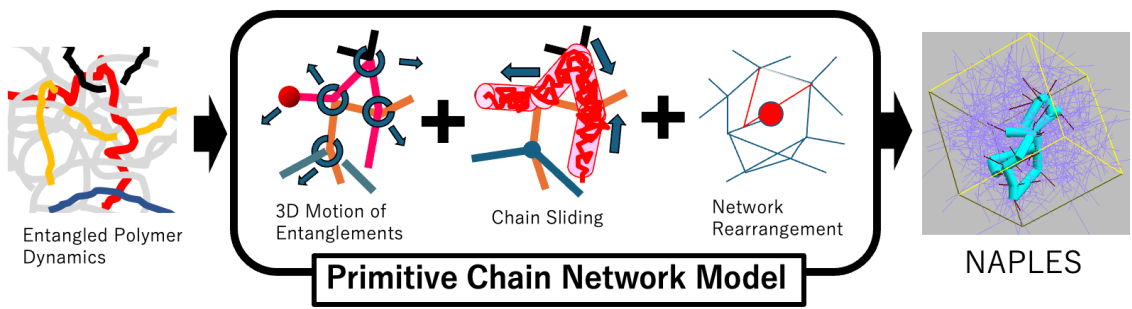

**Figure 1** Schematic representation of the primitive chain network (PCN) model, in which the entangled polymer motion is decomposed into the motion of entanglement points, chain sliding, and network rearrangement. The rightmost panel shows a typical snapshot

of the simulation code, named New Algorithm for Polymeric Liquids Entangled and Strained (NAPLES).

The dynamics are governed by Langevin-type equations. The equation of motion for the slip-link position accounts for the force balance[34] as follows.

$$\mathbf{0} = -2\zeta(\dot{\mathbf{R}} - \mathbf{\kappa} \cdot \mathbf{R}) + 3k_BT \frac{n_0}{a^2} \sum_{j}^{4} \frac{\mathbf{r}_j}{n_j} - n_0 \nabla \mu + \mathbf{F}_B \qquad (1)$$

The first term on the right-hand side is the drag force, where $\zeta$ is the friction coefficient of the single strand, and $\mathbf{\kappa}$ is the velocity gradient tensor of the background flow. Although the friction coefficient is often assumed constant, it may vary under fast and large deformations due to stretch/orientation-induced reduction (SORF)[35–38]. The prefactor of 2 reflects that two entanglement segments are bundled into a single slip-link. $\mathbf{\kappa}$ is controlled via a feedback loop when creep is simulated[39]. The second term is the tension in the strands connected to the node with the average number of Kuhn segments $n_0$, and the average strand length $a$ under equilibrium. According to the binary assumption of entanglement, 4 strands diverge from each node. Although the Gaussian tension is used in this formula, finite extensibility can be implemented as needed[40]. Bending stiffness can also be implemented[41]. The third term is the osmotic force that suppresses density inhomogeneity in the system, and $\mu$ is the chemical potential. The fourth term is the Gaussian random force.

Regarding the osmotic force, the chemical potential is phenomenologically chosen as follows, with the local segment density $\phi$.

$$\frac{\mu}{k_BT} = \begin{cases} \varepsilon \left(\frac{\phi}{\langle \phi \rangle} - 1\right)^2 & \text{for } \phi > \langle \phi \rangle \\ 0 & \phi \leq \langle \phi \rangle \end{cases} \qquad (2)$$

Here, $\langle \phi \rangle$ is the global average of $\phi$, and $\varepsilon$ is the intensity parameter being of order unity. In addition to this field-type implementation, soft-core interactions between network nodes were examined[42].

Coupled with the 3D motion of the nodes, the transport of monomers between adjacent subchains via the slip-links is described by a 1D kinetic equation for $n$, which represents the chain sliding (reptation) and contour length fluctuations (CLF)[43]. The rate of change of monomer number $n_j$ in subchain $j$ is driven by the tension difference between adjacent subchains and the chemical potential gradient along the chain, as described below.

$$0 = -\zeta \frac{\dot{n}}{\rho} + 3k_B T \frac{n_0}{a^2} \left( \frac{r_{j+1}}{n_{j+1}} - \frac{r_j}{n_j} \right) - n_0 \nabla \mu + f_B \qquad (3)$$

Here, $f_B$ is the Gaussian random number.

As a consequence of the sliding dynamics described above, the chain may slide off or protrude from the connected slip-link, inducing rearrangement of the network and fluctuation of $Z$. If $n$ in a chain-end subchain falls below a critical threshold, the slip-link (and the network node) connecting this end subchain is destroyed. When a slip-link is destroyed, the constraint is removed not only from the retracting chain but also from the entangled partner chain, realizing constraint release (CR). Conversely, if $n$ exceeds an upper threshold, a new slip-link (and a network node) is created. The chain end "hooks" a surrounding strand chosen geometrically from a sphere of radius $a$ centered at the middle of the end subchain. These constraint renewals naturally lead to a distribution of subchain lengths and entanglement numbers at equilibrium[44–47]. It has been observed that the equilibrium distribution of monomers in a subchain is not a simple exponential[48] but exhibits a peak, which aligns with theoretical predictions for small systems[49] and blocking effects[50].

For branch polymers, particularly those with multiple branch points along the backbone, entanglements between backbones cannot be relaxed by network reconstruction at the chain ends. In this case, two algorithms are considered: branch point reptation (BPR)[51] and branch point withdrawal (BPW)[52]. BPR is based on the hierarchical relaxation picture[53,54], in which the branch point is allowed to hop along the backbone when the attached arm is fully relaxed. The other mechanism, BPW, is triggered by the tension difference across the backbone[55]; when the tension in the backbone segment exceeds the combined tension of the branching arms, the branch point is sucked into the slip-link of the backbone. BPW rarely occurs under equilibrium conditions, whereas it becomes dominant under rapid, large deformations.

Units of length, energy, and time are chosen as $a$, $k_BT$, and the diffusion time of the node $\tau_{\text{PCN}} = \zeta a^2/6k_BT$. However, for convenience in direct comparison with experimental systems, units of molecular weight and modulus are used instead of $a$ and $k_BT$. Namely, the unit molecular weight $M_{\text{PCN}}$ corresponds to the molar mass of the entanglement segment with the equilibrium length of $a$. The unit modulus is defined as $G_{\text{PCN}} = k_BT/a^3$. These parameters are conceptually similar to, but quantitatively different from, the entanglement molecular weight $M_e$ and the plateau modulus $G_N$, as mentioned later.

A simulation code based on the PCN model has been released under the name NAPLES (New Algorithm for Polymeric Liquids Entangled and Strained)[25], named also after the city in southern Italy where the model was developed.

## 3. Multi-Chain Slip-Spring (MCSS) Model

The PCN model is robust for predicting rheological behavior and capturing many-body effects. However, its phenomenological rules for slip-link creation/destruction and the ad hoc osmotic potential lack a rigorous derivation from a Hamiltonian. The thermodynamic consistency, particularly regarding detailed balance at equilibrium, is not guaranteed[56]. The multi-chain slip-spring (MCSS) model was developed to address these issues, following the single-chain version[57].

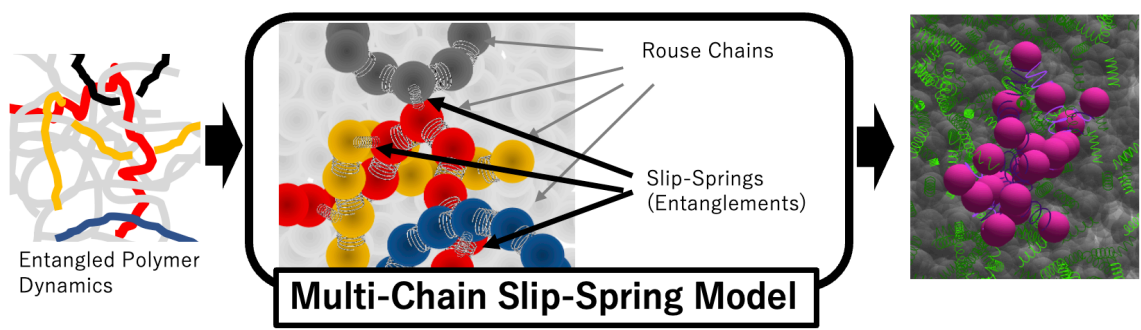

**Figure 2** Schematic representation of the multi-chain slip-spring (MCSS) model, in which the entangled polymer network is represented by Rouse chains connected via slip-springs. The rightmost panel is a typical snapshot of the simulation code.

The MCSS model is based on a well-defined free energy functional[27]. The system consists of Rouse chains (beads connected by harmonic springs) in a simulation box.

Entanglements are represented by virtual "slip-springs" that connect beads on different chains (or beads on the same chain as a knot), as schematically shown in Fig 2. The total free energy $\mathcal{F}$ of the system is explicitly defined as:

$$\frac{\mathcal{F}(\{\mathbf{R}_{i,k}\},\{S_\alpha\},Z)}{k_B T}$$

$$= \sum_{i,k} \frac{3(\mathbf{R}_{i,k+1} - \mathbf{R}_{i,k})^2}{2b^2} + \sum_{\alpha=1}^{Z} \frac{3(\mathbf{R}_{S_{\alpha,1},S_{\alpha,2}} - \mathbf{R}_{S_{\alpha,3},S_{\alpha,4}})^2}{2N_s b^2}$$

$$+ e^{\nu/k_B T} \sum_{i,k,j,l} \exp\left[-\frac{3(\mathbf{R}_{i,k} - \mathbf{R}_{j,l})^2}{2N_s b^2}\right] \quad (4)$$

The first term in the right-hand side is the standard entropic elasticity of the Rouse chains, where $\mathbf{R}_{i,k}$ is the position of bead $k$ on chain $i$, and $b$ is the average spring length. The second term is the elastic energy due to additional slip-springs, for which $N_s$ is a parameter controlling the slip-spring intensity, and $S_{\alpha,j}$ is the anchoring point of slip-spring $\alpha$. The third term is a repulsive potential introduced to compensate for the artificial attraction induced by the slip-springs. This term exactly recovers the Gaussian chain statistics, whereas the soft-core interaction used in dissipative particle dynamics (DPD) simulations is sufficient to eliminate the influence of the second term on the chain conformation[58–61].

The dynamics are derived to satisfy the detailed balance condition associated with the probability distribution function $P$ related to the free energy as $P \propto \exp(-\mathcal{F}/k_B T)$. The bead motion is described by the Fokker-Planck equation.

$$\frac{\partial P}{\partial t} = \sum \frac{1}{\varsigma} \frac{\partial}{\partial R}\left[\frac{\partial \mathcal{F}}{\partial R}P + k_B T \frac{\partial P}{\partial R}\right] \quad (5)$$

Here, $\varsigma$ is the friction constant. The connectivity matrix evolves according to the master equations for the transition rate $W$ of the connectivity $S$ and the number of slip-springs on each chain $Z$.

$$\frac{\partial P}{\partial t} = W(S|S')P(S') - W(S'|S)P(S) \quad (6)$$

$$\frac{\partial P}{\partial t} = W(Z|Z')P(Z') - W(Z'|Z)P(Z) \tag{7}$$

The numerical implementation is achieved via a Metropolis or Glauber Monte Carlo scheme. The rates $W$ are designed such that the instantaneous number of slip-springs fluctuates around an equilibrium value determined by the fugacity $e^{\nu/k_B T}$.

The original MCSS model[27] allows multiple slip-springs to share a single bead. A variant was also examined to eliminate this "degeneracy" of slip-springs by restricting sites to hold at most one slip-spring[62]. This introduces Fermi-Dirac rather than Bose-Einstein statistics for the slip-springs, but the core principle of deriving dynamics from a defined Hamiltonian remains the same. The Fermi-Dirac variant, however, does not work properly due to an asymmetric exclusion process, where the sliding motion of slip-links is significantly retarded, like motor cars in a traffic jam.

Units of length, energy, and time are the bond length $b$, thermal energy $k_B T$, and the bead diffusion time $\tau_{\text{MCSS}} = \varsigma b^2 / 6 k_B T$. Similar to the case of PCN, for comparison to real systems, the unit molecular weight $M_{\text{MCSS}}$ corresponds to the molar mass of the single bead and the unit modulus $G_{\text{MCSS}}$ are used, instead of $b$ and $k_B T$. Note that $G_{MCSS}$ is not written as $k_B T/b^3$ but related to the number density of slip-springs, as discussed below.

### 4. Plateau Modulus and Entanglement Molecular Weight

The relationship between the plateau modulus $G_N^0$ and the entanglement molecular weight $M_e$ is frequently written as $G_N^0 = \rho RT/M_e$. Note that this relation depends on the theoretical assumption that links the modulus to the characteristic molecular weight[45], and a convenient form is $G_N^0 = A\rho RT/M_0$, with model-dependent parameters $A$ and $M_0$. Since $G_N^0$ is an experimentally-defined quantity, $M_0 = AM_e$. For instance, the phantom network and tube theories give $A = 1/2$ and $4/5$, implying that $M_0 = M_e/2$ and $4M_e/5$, reflecting thermal fluctuations in the network node positions in 3D and in the chain sliding along the chain contour. The PCN model has both fluctuations[63,64] and $A = 4/7$, consistent with the slip-tube model by Rubinstein and Panyukov[65,66]. In the MCSS model, in addition to the fluctuations implemented in the PCN model, the entanglement nodes exhibit internal fluctuations. According to the single-chain theory by Uneyama[46], the $A$ value depends on the parameters in the MCSS model as follows.

$$A = \frac{11}{15}\left\{1 + 4\left(\frac{N_s}{N_{eSS}^\infty}\right)\right\}^{-\frac{1}{2}} \tag{8}$$

$$\frac{1}{N_{eSS}^\infty} \approx 2e^{\nu/k_BT}\rho_b\left(\frac{2\pi N_s b^2}{3}\right)^{\frac{3}{2}} \tag{9}$$

Here, $\rho_b$ is the Rouse bead number density, and $N_{eSS}^\infty$ is the average Rouse bead number between consecutive anchoring points of slip-springs along the Rouse chain. Although these equations provide a theoretical guide, the value of $A$ is not fully consistent with the simulation results, likely due to differences between the single-chain modeling in the theory and the multi-chain simulations. For instance, with typical parameters chosen at $e^{\nu/k_BT} = 0.036$, $\rho_b = 4$, and $N_s = 0.5$, eqs 8 and 9 give $N_{eSS}^\infty \sim 3.2$ and $A \sim 0.58$. Although this $N_{eSS}^\infty$ value is consistent with simulations, $A$ value used for the fitting of simulation results to experimental data is ca. 0.2.

5. Mapping and Universality

Since PCN and MCSS models reproduce the universal behavior of entangled polymer dynamics, their results are compatible with each other, and with microscopic simulations, when converted using conversion factors for molecular weight, length, time, and modulus[67–69]. (Note that modulus is considered instead of energy for convenience.) For the standard bead-spring model proposed by Kremer and Grest, the conversion can be achieved according to the following relationships.

$$N_{KG} = 30N_{MCSS}/N_{eSS}^\infty = 40Z_{PCN} \tag{10}$$

$$\sigma_{KG}^2 = 35b^2/N_{eSS}^\infty = 38a^2 \tag{11}$$

$$\tau_{KG} = 1.5 \times 10^4 \tau_{MCSS}/{N_{eSS}^\infty}^2 = 1.3 \times 10^4 \tau_{PCN} \tag{12}$$

$$G_{KG} = 2.2 \times 10^{-2} G_{MCSS}(N_{eSS}^\infty/A\rho_b) = 1.9 \times 10^{-2} G_{PCN} \tag{13}$$

Here, subscripts denote KG (Kremer-Grest), MCSS, and PCN models, respectively. Regarding KG and PCN models, each model has a fixed level of coarse-graining; thus, a single set of conversion relations is established. On the contrary, the level of coarse-graining is arbitrarily chosen for MCSS through $N_{\text{eSS}}^{\infty}$ by the choice of parameters[60,70]. See eq 9. This feature is inherited from the Rouse model.

Figure 3 shows the mapping to the results of the KG simulations for the diffusion and relaxation moduli[67]. With the conversion factors given by eqs 10-13, the KG results are quantitatively reproduced by MCSS and PCN. Although not shown here, the MCSS simulations with different $N_{\text{eSS}}^{\infty}$ values overlap with each other, with shifts according to the conversion factors[70]. The DPD-SS models require a different set of conversion factors[60].

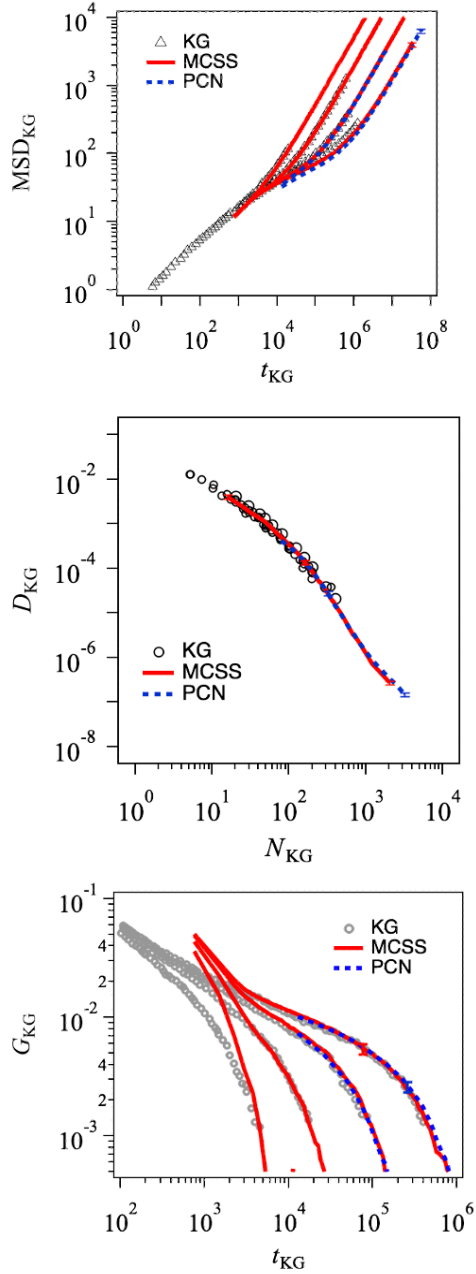

**Figure 3** Comparison of MSD (top), diffusion constant (middle), and relaxation modulus (bottom) among Kremer-Grest (symbols), MCSS (red curves), and PCN (blue broken curves) simulations, with conversions according to eqs 10-13. The MCSS results are for the case with $N_{eSS}^{\infty} = 3.3$. For the Kremer-Grest simulations in the top and bottom panels, $N_{KG}$ =50, 100, 200, and 350, from left to right, respectively[67].

Figure 4 shows a comparison of the dynamic viscoelasticity and zero-shear viscosity of H-branch polystyrene melt between experimental data[71] and simulation results[69]. Since the level of coarse-graining differs, the covered frequency range differs between PCN and MCSS. Nevertheless, both simulation results quantitatively capture the viscoelastic relaxation. Consequently, the molecular-weight dependence of viscosity is well reproduced.

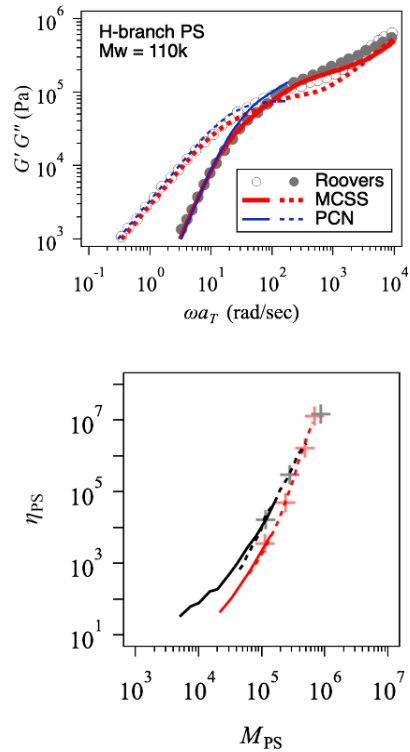

**Figure 4** Comparison of dynamic viscoelasticity (top) and zero-shear viscosity as a function of molecular weight (bottom) for H-branch polystyrene melt between experimental data and the simulation results for PCN and MCSS[69]. In the bottom panel, the data for linear and H polymers are shown in black and red, respectively. The experimental data[71] are indicated by symbols, whereas MCSS and PCN results are shown as solid and broken curves.

Figure 5 shows the normalized flow curve[72] compared with an experimental result[73]. A reasonable agreement is found for the shear-thinning behavior if the molecular weight is adequately mapped. Concerning MCSS, the result is insensitive to the value of $N_{eSS}^{\infty}$, similar to the linear viscoelasticity.

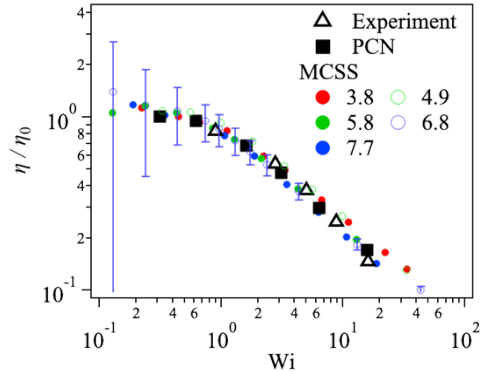

**Figure 5** Normalized flow curves for an entangled polyisoprene solution with M/Me ~ 24 (triangle) compared with PCN (square) and MCSS (circle) results[72]. For MCSS, various $N_{eSS}^{\infty}$ values indicated in the panel were examined and exhibited universal behavior.

## 6. Applications

In general, the primary purpose of molecular simulations is twofold: to predict material properties and to analyze molecular mechanisms.

For the first direction, PCN simulations have been used owing to their greater coarse-graining and lower computational cost than MCSS. A typical example shown in Fig 6 is the combination of PCN with continuum-flow simulations in polymer processing[74]. Residual stress in a thin-walled injection-molded product was predicted using linear and nonlinear rheology obtained from PCN simulations, and the effects of molecular weight on warpage were analyzed. In another example of the reverse, PCN simulations were performed to observe molecular motion during melt spinning, based on the flow field calculated by the continuum model.

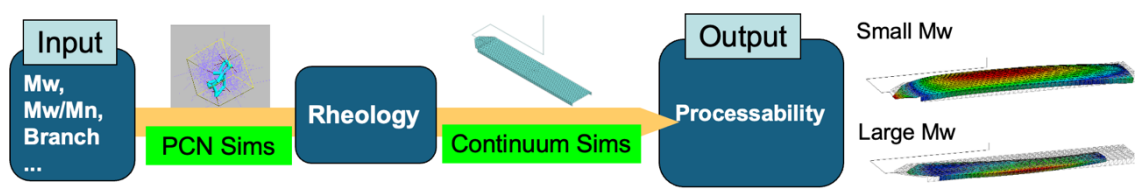

**Figure 6** Schematic of a multi-scale simulation of polymer processing combining PCN and continuum flow simulations to predict warpage of thin-walled injection-molded products from the molecular weight of the material [74].

In the second direction, linear and nonlinear rheology was analyzed in terms of contributions from segment orientation, chain stretch, entanglement density, and other factors. For instance, aligning the multi-chain construction with force balance around entanglements, statistics of entanglement network structure, and their relation to plateau modulus were discussed[45]. The time-stress discrepancy, the inconsistency in the predictions for relaxation time and plateau modulus in Doi's contour-length fluctuation (CLF) theory, was also examined, and the effects of constraint release (CR) were revealed[75]. The effects of CR on the dilution exponent[76] and the relation to the tube survival fraction[77] were investigated separately. In relation to CR, correlations between different chains[78–81] contribute non-negligibly (ca. 50%) and are time-dependent in the relaxation modulus in both PCN and MCSS (see Fig 7)[67], posing a fundamental difficulty for single-chain modeling. As a basic test of the implementation of CR, bidispese systems were also examined[82,83]. For branch polymers, the effects of BPR on relaxation modulus and diffusion were also quantitatively demonstrated[69,84].

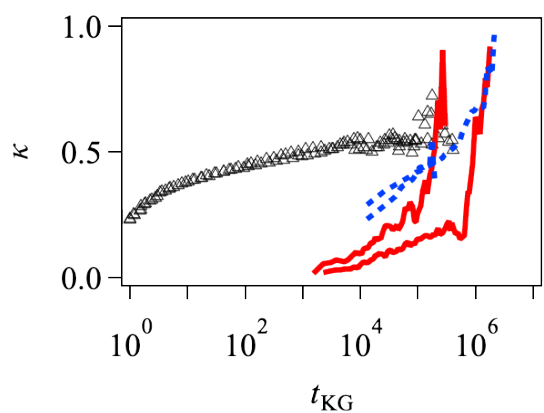

**Figure 7** Contribution from the cross-correlation between different chains in the linear relaxation modulus for $N_{KG} = 200$ and 350 chains. Symbols, red solid curves, and blue broken curves correspond to KG, MCSS, and PCN simulations, respectively[67].

Regarding nonlinear rheology, the basic assumption of the Doi-Edwards theory for large-step deformations was examined; changes in entanglement density and the contribution of chain retraction were observed for PCN[85–90]. Under start-up shear deformations, the mechanisms of stress overshoot[91–94] (including double overshoots[95,96]) and undershoot[97–99] were clarified in both PCN and MCSS simulations in terms of segment orientation, stretch, molecular tumbling, and conformational relaxations[100–104]. Analysis of interrupted shear flows was also performed[105]. For elongational flows, as well as the effects of finite chain extensibility[40], changes in the friction constant have been thoroughly examined[36,106–111]; the core idea is that in a melt, the monomeric friction coefficient is not constant but depends on the degree of orientation of the surrounding chains, and it reduces to 1/10 from the equilibrium value, as shown in Fig 8. By introducing this mechanism, the difference between melts and entangled solutions under fast elongation can be quantitatively reproduced. The effects of finite chain extensibility[40] and BPW[52,92,108] were also quantitatively investigated.

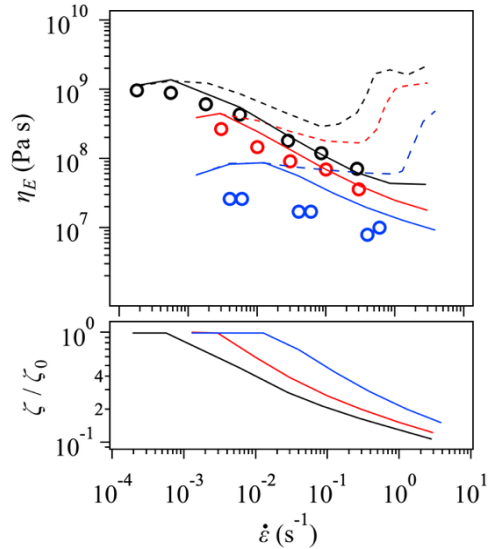

**Figure 8** (Top) Uniaxial elongational flow-curves for nearly-monodispersed polypropylene carbonate with the molecular weights of 69k (blue), 111k (red), and 158k (black). Symbols are experimental data[112]. PCN simulation results[106,107] with and without friction reduction are drawn by solid and broken curves. (Bottom) Normalized friction coefficients to the equilibrium values $\zeta_0$ in the elongation simulations.

## 7. Model Extensions

Owing to the real-space, multi-chain nature of the models, several directions have been explored for extending them to fancy problems. For the PCN model, slip boundary conditions at solid walls[113–115] were implemented, as shown in Fig 9, and the effect of slippage on macroscopic rheology was discussed. Non-isotropic orientational relaxation in confined geometries was also observed. The other direction is gelation, in which a fraction of slip-links is converted into cross-links[116] or cross-linkers[117] are introduced, as shown in Fig 10. The gelation kinetics and the mechanical properties of resultant networks were investigated, and the role of trapped entanglements was discussed.

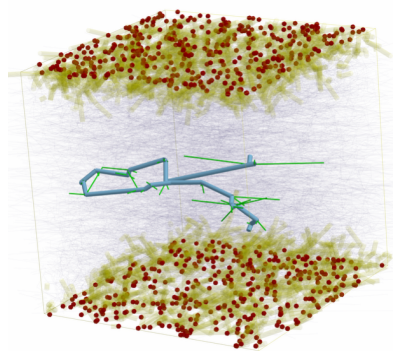

**Figure 9** A typical snapshot of the extended PCN model implementing solid walls with grafted chains under shear deformation.

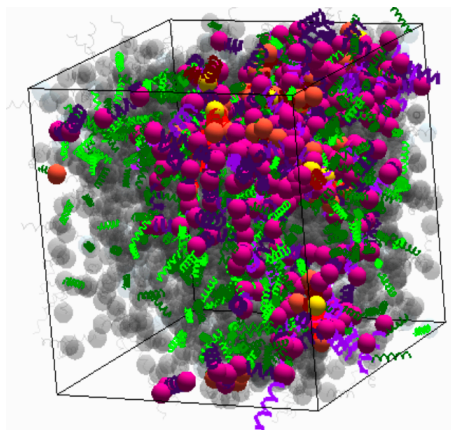

**Figure 10** A percolated network with trapped entanglements (green springs) obtained from the extended MCSS simulation with cross-linkers.

## 8. Conclusion and Outlook

Over the past two decades, multi-chain models for entangled polymer dynamics have developed from the phenomenological PCN model to the thermodynamically rigorous MCSS model. In the PCN model, force balance at entanglement nodes is explicitly resolved in three-dimensional space. This construction naturally gives rise to constraint release, convective constraint release, and hierarchical relaxation effects in branched polymers, without ad hoc parameters. The PCN model also served as a platform for testing hypotheses such as friction reduction under fast flows. The MCSS model placed the simulation on a firm statistical mechanical basis. All dynamics, including the hopping and creation/destruction of slip-springs, are derived from a well-defined free energy, ensuring thermodynamic consistency and detailed balance. The mapping rules for KG models established PCN and MCSS as quantitative tools that connect atomistic/bead-spring MD and macroscopic continuum mechanics.

Several directions remain for further development. The most important are inhomogeneous systems, including blends[85], copolymers[118,119], vitrimers, and composites[120]. PCN has been applied to some of these systems, but it lacks thermodynamic rigor. MCSS and its DPD version are thermodynamically capable, but the kinetics of slip-springs in inhomogeneous systems are non-trivial. Another challenging direction is crystallization, which may be modeled using slip springs, as suggested by recent work of Uneyama[121]. A different direction altogether is the use of these high-throughput simulations to assist machine learning and AI. Studies along these lines are ongoing.

**Acknowledgements**

The author thanks G. Marrucci, G. Ianniruberto, F. Greco, H. Watanabe, T. Uneyama, and F. Müller-Plathe for their contributions to the work. The financial support from JSPS, NEDO, JST, and private agencies and companies is also acknowledged.